\documentstyle[prl,aps]{revtex}
\begin{document}
\twocolumn
\draft       
\widetext

\title{\bf Reply to the Comment of R. M. Cavalcanti on
``Resonant Spectra and the Time
Evolution of the Survival and Nonescape Probabilities''}
\maketitle
\narrowtext

In our paper\cite{prl} we derived an exact expression for the
nonescape probability $P(t)$, (see Eq. (14)), as an
expansion in terms of resonant states and {\it M} functions,
\begin{equation}
P(t)= \sum_{n=-\infty}^{\infty} \sum_{\ell=-\infty}^{\infty}
C_n C^{\ast}_{\ell}I_{n \ell}
M (k_n,t) M^{\ast}(k_{\ell},t),
\label{1}
\end{equation}
where the integral $I_{n \ell}$ is defined by Eq. (15) of ref. \cite{prl},
\begin{equation}
I_{n \ell} = \int^R_0  u^*_{\ell}(r) u_n(r) dr.
\label{3}
\end{equation}
\noindent
The long time limit of $P(t)$ leads to an asymptotic
expansion whose leading term reads,
\begin{equation}
\sum_{n=-\infty}^{\infty}\sum_{\ell=-\infty}^{\infty}
\left( \frac {C_nC_{\ell}^* I_{n \ell}}{k_n k_{\ell}^*} \right )
{1 \over t}.
\label{2}
\end{equation}
So we concluded that at long times $P(t) \sim t^{-1}$. Cavalcanti\cite{com}
instead has proven that the above coefficient vanishes
and concludes that the leading term of $P(t) \sim t^{-3}$.
His procedure corresponds to interchange the integral over $r$ in the
expression of $P(t)$, Eq.\ (\ref{1}), with the long time limit.
The vanishing of the term proportional to $t^{-1}$ then follows from
the sum rule
\begin{equation}
\sum_{m=-\infty}^{\infty} \frac {C_mu_m(r)}{k_m}=0,\,\,\,\,\,(r \leq R).
\label{2a}
\end{equation}

In our approach we perform first the integration over $r$ and
then take the long time limit.
We provide below an argument that shows that in dealing with resonant state
expansions the interchange of the above operations do not lead to the same
result.
This is the case for expansions that do not converge uniformly.

Consider the {\it n-th} resonant function $u_n(r)$ is a solution of the
Schr\"odinger equation\cite{ggc},
\begin{equation}
u^{''}_n(r) + [k_n^2-V(r)]u_n(r)=0,
\label{4}
\end{equation}
where the prime stands for the derivative with repect to $r$,
$k_n^2$ is a squared complex wavenumber, and $V(r)$
is an arbitrary potential
that vanishes beyond $r=R$. The function $u_n(r)$ satisfies the
boundary conditions,
\begin{equation}
u_n(0)=0;\,\,\,\,\,\, \left [u^{'}_n(r) \right ]_{r=R}=ik_n u_n(R).
\label{5}
\end{equation}
Consider also similar equations for the complex $\ell-th$ function
$u_{\ell}^*(r)$,
\begin{equation}
u^{''*}_{\ell}(r) + [k^{*2}_{\ell}-V(r)]u_{\ell}^*(r)=0,
\label{6}
\end{equation}
which obeys the boundary conditions,
\begin{equation}
u^*_{\ell}(0)=0;\,\,\,\,\,\,
\left [u^{'*}_{\ell}(r)\right ]_{r=R}=-ik_{\ell}^{*}u^{*}_{\ell}(R).
\label{7}
\end{equation}
Now multiply Eq.\ (\ref{4}) by $u_{\ell}^*(r)$ and substract from
it Eq.\ (\ref{6})
multiplied by $u_n(r)$. Integrating the resulting expression from
$r=0$ to $r=R$ yields,
\begin{equation}
\left [u_n(r)u_{\ell}^{'*}(r)- u_{\ell}^{*}(r)u_n^{'}(r)
\right ]_{r=0}^{r=R}
+(k_n^2-k_{\ell}^{*2})I_{n \ell}=0.
\label{8}
\end{equation}
Using Eqs.\ (\ref{5}) and (\ref{7}) allows to write a closed form
of $I_{n \ell}$,
\begin{equation}
I_{nl}= {u_n(R)u_{\ell}^*(R)\over i(k_n-k_{\ell}^*)}.
\label{9}
\end{equation}
Substituting Eq.\ (\ref{9}) into Eq.\ (\ref{1}) leads to the
following exact expression for the nonescape probability,
\begin{equation}
P(t)= \sum_{n=-\infty}^{\infty} \sum_{\ell=-\infty}^{\infty}
C_n C^{\ast}_{\ell} {u_n(R)u_{\ell}^*(R) \over i(k_n-k_{\ell}^*)}
M (k_n,t) M^{\ast}(k_{\ell},t).
\label{10}
\end{equation}
Taking now the long time limit allows to write  $P(t)$ at leading
order in inverse powers of $t$ as,
\begin{equation}
P(t) \sim  \sum_{n=-\infty}^{\infty} \sum_{\ell=-\infty}^{\infty}
\left (  {C_n \over k_n} {C^{\ast}_{\ell} \over k_{\ell}^*}
{u_n(R)u_{\ell}^*(R)
\over i(k_n-k_{\ell}^*)} \right ){1 \over t}.
\label{11}
\end{equation}
The sum rule given by Eq.\ (\ref{2a}) does not lead to the vanishing of
Eq.\ (\ref{11}) because of the existence of the factor $1/(k_n-k_{\ell}^*)$.
This shows that the interchange of integration and the long time
limit operations on the resonant expansions yields different results.
In our opinion, according to the definition of $P(t)$, the integration
over $r$ should precede the long time limit and consequently
$P(t) \sim t^{-1}$.
\vskip1truecm
\noindent
G. Garc\'{\i}a-Calder\'on, J.L. Mateos,  M. Moshinsky

\noindent
Instituto de F\'{\i}sica

\noindent
Universidad Nacional Aut\'onoma de M\'exico

\noindent
Apartado Postal 20-364

\noindent
01000 M\'exico, D.F.

\noindent
M\'exico


\begin{references}
\bibitem{prl} G. Garc\'{\i}a-Calder\'on, J. L. Mateos and
M. Moshinsky, Phys. Rev. Lett. {\bf 74}, 337 (1995).
\bibitem{com} R. M. Cavalcanti, Phys. Rev. Lett. preceding Comment.
\bibitem{ggc} G. Garc\'{\i}a-Calder\'on and R.E. Peierls, Nucl. Phys.
{\bf A 265}, 441 (1976); G. Garc\'{\i}a-Calder\'on, J. L. Mateos and
M. Moshinsky, Ann. Phys. (N.Y.) {\bf 249}, 430 (1996).
\end{references}
\end{document}